\documentclass[aps,showpacs,nofootinbib,preprintnumbers,amsmath,amssymb]{revtex4}

\usepackage{graphicx}
\usepackage{dcolumn}
\usepackage{bm}
\usepackage{epsfig}
\usepackage[usenames]{color}
\usepackage{hyperref} 

\newcommand{\stackeven}[2]{{{}_{\displaystyle{#1}}\atop\displaystyle{#2}}}

\newcommand{\as}{\alpha_s}

\newcommand{\bas}{{\bar\alpha}_s}
\newcommand{\bam}{{\bar\alpha}_\mu}

\def\eq#1{{Eq.~(\ref{#1})}}

\newcommand{\ben}{\begin{eqnarray*}}
\newcommand{\een}{\end{eqnarray*}}

\newcommand{\amu}{\alpha_\mu}
\newcommand{\pd}{\partial}

\newcommand{\dhd}{{\textstyle d}
\lower.03ex\hbox{\kern-0.38em$^{\scriptstyle-}$}\kern-0.05em{}}
\newcommand{\dbar}{{\textstyle \delta}
\lower.03ex\hbox{\kern-0.38em$^{\scriptstyle-}$}\kern-0.05em{}}

\begin{document}

\title{Solution of the NLO BFKL Equation \\ and a Strategy for Solving
  the All-Order BFKL Equation}

\author{Giovanni~A.~Chirilli,\footnote{chirilli.1@asc.ohio-state.edu}
  Yuri~V.~Kovchegov\footnote{kovchegov.1@asc.ohio-state.edu}}

\affiliation{Department of Physics, The Ohio State University,
  Columbus, OH 43210, USA}

\begin{abstract}
  We derive the solution of the NLO BFKL equation by constructing its
  eigenfunctions perturbatively, using an expansion around the LO BFKL
  (conformal) eigenfunctions. This method can be used to construct a
  solution of the BFKL equation with the kernel calculated to an
  arbitrary order in the coupling constant. 
\end{abstract}


\maketitle



\section{Introduction}

At very high energy the dynamics of hadronic and nuclear scattering
processes is dominated by non-linear effects associated with parton
saturation (see
\cite{Jalilian-Marian:2005jf,Weigert:2005us,Iancu:2003xm,Gelis:2010nm,KovchegovLevin}
for the reviews of the subject). Scattering amplitudes of these
processes can be factorized in rapidity-space using the operator
product expansion in terms of Wilson-lines, infinite gauge links
ordered along the straight line collinear to the particle's velocity
close to the light-cone \cite{Balitsky:2001gj}. The evolution of
Wilson-lines with respect to rapidity is given by the non-linear
Balitsky--Kovchegov (BK)
\cite{Balitsky:1996ub,Balitsky:1998ya,Kovchegov:1999yj,Kovchegov:1999ua}
and Jalilian-Marian--Iancu--McLerran--Weigert--Leonidov--Kovner
(JIMWLK)
\cite{Jalilian-Marian:1997dw,Jalilian-Marian:1997gr,Iancu:2001ad,Iancu:2000hn}
evolution equations.  The pre-asymptotic behavior of the scattering
amplitude, when the parton density is not yet high enough for
saturation effects to become important, is well-described by the
linear Balitsky--Fadin--Kuraev--Lipatov (BFKL)
\cite{Kuraev:1977fs,Kuraev:1976ge,Balitsky:1978ic} evolution equation.

Over the last decade the physics of saturation/Color Glass Condensate
(CGC) achieved a number of phenomenological successes, in both
predicting and describing the experimental data (see
e.g. \cite{Albacete:2010sy,ALbacete:2010ad}). Current phenomenological
applications mainly employ the BK evolution in the leading logarithmic
approximation (LLA), which resums powers of $\as \, \ln s$, with the
running coupling corrections included in the evolution kernel
\cite{Balitsky:2006wa,Gardi:2006rp,Kovchegov:2006vj,Kovchegov:2006wf}. (Here
$\as$ is the strong coupling and $s$ is the center-of-mass energy.)
Further progress in terms of precision of the CGC predictions and in
understanding the accuracy of the CGC fits can be achieved by studying
other higher-order corrections to the BFKL/BK/JIMWLK evolution
equations. It is, therefore, very important to better understand the
structure of high-order corrections to the small-$x$ evolution
equations.

At the moment, the full next-to-leading-order (NLO) kernel is known
for the BFKL \cite{Fadin:1998py,Ciafaloni:1998gs} and BK
\cite{Balitsky:2008zz} equations, though phenomenological
implementation of the latter is presently lacking. The calculation of
the NLO BFKL kernel some fifteen years ago
\cite{Fadin:1998py,Ciafaloni:1998gs} spurred a vigorous activity
trying to understand the effects of NLO corrections on the solution of
the leading-order (LO, or LLA) BFKL equation.

The solution of the LO BFKL evolution
\cite{Kuraev:1977fs,Kuraev:1976ge,Balitsky:1978ic} describes a scattering
amplitude that grows proportionally to a positive power of the center
of mass energy of the hadronic or nuclear scattering process: at this
order the kernel of the evolution equation respects the conformal
symmetry of the $SL(2,C)$ M\"{o}bius group and the eigenfunctions are
power-like functions of transverse distance in coordinate space (or,
in momentum space, powers of transverse momenta), while the eigenvalue
of the kernel is related to the Pomeron intercept. At the NLO there
appears also a contribution to the evolution kernel due to the running
of the QCD coupling constant and the conformal property of the LO BFKL
is lost. Consequently, the LO BFKL kernel's conformal eigenfunctions
are not eigenfunctions of the NLO BFKL kernel: at this order the
power-law growth of the amplitudes with energy also seems to be lost
because of the non-Regge terms appearing due to the running coupling
effects
\cite{Kovchegov:1998ae,Levin:1998pka,Armesto:1998gt,Ciafaloni:2001db}.
It was also found that the NLO BFKL corrections generate a large
negative contribution to the eigenvalue/Pomeron intercept
\cite{Fadin:1998py,Ross:1998xw,Andersen:2003an,Andersen:2003wy},
though resummations of collinear divergences to all higher orders in
the kernel offer a promising way to tame this large correction
\cite{Salam:1998tj,Ciafaloni:1999yw,Ciafaloni:2003rd}.

Despite a number of efforts, it appears that an exact analytical
solution of the NLO BFKL equation, or a general all-order BFKL
equation in QCD is still lacking (see
e.g. \cite{Bondarenko:2008uc}). This is in stark contrast to the
Dokshitzer--Gribov--Lipatov--Altarelli-Parisi (DGLAP) evolution
equation \cite{Dokshitzer:1977sg,Gribov:1972ri,Altarelli:1977zs},
which is a renormalization group equation in the virtuality $Q^2$: the
eigenfunctions of that evolution equation are simple powers of
Bjorken-$x$ variable for the kernel calculated to any order in the
coupling constant. The general form of the solution for DGLAP equation
is well-known, with the higher-order corrections in the powers of the
coupling constant entering into the anomalous dimension of the
operator at hand
\cite{Gross:1973id,Georgi:1951sr,Altarelli:1977zs,Floratos:1977au,Vogt:2004mw}.

The goal of this work is to devise a systematic way of solving the
BFKL evolution equation order-by-order in the perturbation theory. We
consider scattering in the pre-asymptotic regime, where the non-linear
saturation corrections due to BK/JIMWLK evolution are not yet
important, and the linear BFKL equation gives a good description of
the scattering amplitude.

Note that in the ${\cal N} =4$ super-Yang-Mills (SYM) theory, which is
a conformal field theory without the running of the coupling, the form
of the solution of the all-order BFKL equation is known
\cite{Kotikov:2000pm,Balitsky:2008rc,Balitsky:2009xg,Balitsky:2009yp}:
conformal symmetry fixes the eigenfunctions of the BFKL kernel in
${\cal N} =4$ SYM theory to be the eigenfunctions of the Casimir
operators of the M\"{o}bius group, $E^{n, \nu}$ \cite{Lipatov:1985uk},
and the perturbative expansion is confined to the eigenvalues/the
intercept.

In QCD the situation is not so straightforward, since, as we have
already mentioned, running coupling corrections destroy the conformal
symmetry of the BFKL kernel, such that the eigenfunctions of NLO (and
higher-order) BFKL equation are not known. Below we will construct the
eigenfunctions of the NLO BFKL by a perturbative expansion in powers
of $\as$ around the conformal eigenfunctions of the LO BFKL
kernel. Naturally the corrections we find are proportional to the QCD
beta-function. Knowing the eigenfunctions and eigenvalues allows us to
find the exact solution of the NLO BFKL equation, given by \eq{NLOsol}
below. (For simplicity we are working in the limit where the
amplitudes do not depend on the azimuthal angles of the transverse
momenta, which corresponds to the dominant high-energy asymptotics.)
Our procedure can be applied to any higher-order BFKL kernel, once it
is calculated. We thus see that the general solution of the BFKL
equation in QCD using our technique contains a perturbative expansion
both in the eigenfunctions and in the eigenvalues, which give the BFKL
Pomeron intercept. This is our proposal for organizing the
perturbation series for the solution of the BFKL equation.

The paper is structured as follows: in Sec. \ref{sec:problem} we
formulate the problem of finding the all-order BFKL Green
function. The NLO eigenfunctions and eigenvalues are found in
Sec.~\ref{sec:NLOsol} and are given in Eqs.~\eqref{eigenf_nu_final}
and \eqref{eigenv3} correspondingly. We see that the eigenvalues we
found are the same as that commonly used in the literature (see
e.g. \cite{Ross:1998xw}). The solution for the NLO BFKL equation is
given in \eq{NLOsol}. The general form of the solution for the
all-order BFKL equation is given in Sec.~\ref{sec:genform} by
\eq{all_sol}. The properties of our solution of the NLO BFKL equation
are studied in Sec.~\ref{sec:NNLOans}, in which we demonstrate the
renormalization scale independence of the obtained solution (within
the perturbative precision). We also rewrite the obtained NLO BFKL
solution in a very compact way in \eq{NLOsol4}. Agreement with the NLO
DGLAP anomalous dimension is established in Sec.~\ref{sec:anomdim}. We
conclude in Sec.~\ref{sec:outlook}.


\section{The Problem: General Form of the Solution of the All-Order
  BFKL equation}
\label{sec:problem}

Our goal is to find the general form of the solution of the
arbitrary-order BFKL equation
\begin{align}
  \label{eq:BFKL}
  \partial_Y G (k, k', Y) = \int d^2 q \, K (k, q) \, G (q, k', Y)
\end{align}
for the Green function $G (k, k', Y)$ with the initial condition
\begin{align}
  \label{eq:init}
  G (k , k' , Y=0) = \frac{1}{2 \pi k} \delta (k-k').
\end{align}
Here $k \equiv |{\vec k}_\perp|$ and $k' \equiv |{\vec k}'_\perp|$ are
the transverse momenta at the two ends of the BFKL ladder, $Y = \ln
(s/ k k')$ is rapidity, $s$ is the center-of-mass energy, and
$\partial_Y \equiv \partial/\partial Y$. For simplicity we consider
only the azimuthally symmetric case, in which the Green function $G$
depends only on magnitudes $k$ and $k'$ of the two-dimensional
transverse vectors ${\vec k}_\perp$ and ${\vec k}'_\perp$: our
technique can be generalized to the case with non-trivial
azimuthal-angle dependence of the scattering amplitude.

Below we will first explicitly solve the NLO BFKL equation, and then
construct the general form of the solution of the all-order BFKL
equation.


\section{Solution of the NLO BFKL equation}

\label{sec:NLOsol}

The kernel $K (k, q)$ of the general BFKL equation \eqref{eq:BFKL} is
known up to the next-to-leading order (NLO)
\cite{Kuraev:1977fs,Kuraev:1976ge,Balitsky:1978ic,Fadin:1998py,Ciafaloni:1998gs}
in the coupling constant. The general BFKL kernel can be written as
\begin{align}
  \label{eq:kernel}
  K (k,q) = \bar{\alpha}_\mu \, K^{\rm LO} (k,q) + \bar{\alpha}_\mu^2
  \, K^{\rm NLO} (k,q) + {\cal O} (\bam^3)
\end{align}
with
\begin{align}
  \label{eq:bam}
  \bam \equiv {\alpha_\mu \, N_c \over \pi}.
\end{align}
(Here $\amu$ is the renormalized strong coupling constant at an
arbitrary renormalization scale $\mu$.) 

The exact form of the leading-order (LO) and NLO kernels $K^{\rm LO}$
and $K^{\rm NLO}$ will not be needed below (see
\cite{Kuraev:1977fs,Kuraev:1976ge,Balitsky:1978ic} and
\cite{Fadin:1998py,Ciafaloni:1998gs} correspondingly). However, it is
essential for us to know the action of the LO and NLO BFKL kernels on
the eigenfunctions of the LO kernel
\cite{Fadin:1998py,Ciafaloni:1998gs}:
\begin{align}
  \int d^2 q \, K^{{\rm LO}+ {\rm NLO}}(k,q) \ q^{2\gamma-2} = \left[
    \bar{\alpha}_\mu \, \chi_0 (\gamma) - \bar{\alpha}^2_\mu \,
    \beta_2 \, \chi_0(\gamma) \, \ln{k^2\over \mu^2} +
    \bar{\alpha}_\mu^2 \, \frac{\delta(\gamma)}{4} \right] \,
  k^{2\gamma-2}
\label{conf-proj}
\end{align}
where
\begin{align}
  \label{eq:LONLO}
  K^{{\rm LO}+ {\rm NLO}}(k,q) \equiv \bar{\alpha}_\mu \, K^{\rm LO}
  (k,q) + \bar{\alpha}_\mu^2 \, K^{\rm NLO} (k,q)
\end{align}
is a shorthand notation for the sum of the LO and NLO kernels. In
\eq{conf-proj} we employed the eigenvalue of the LO BFKL kernel
\begin{align}
  \label{eq:LOeig}
  \chi_0(\gamma)= 2 \, \psi(1) - \psi(\gamma) - \psi(1-\gamma)
\end{align}
with $\psi (\gamma) = \Gamma' (\gamma)/\Gamma (\gamma)$. The
coefficient $\beta_2$ determines the one-loop QCD beta-function
\cite{Gross:1973id,Politzer:1973fx}, such that the corresponding
one-loop QCD running coupling constant is given by
\begin{align}
  \label{eq:beta2}
  \bas (Q^2) = \frac{\bam}{1 + \bam \, \beta_2 \, \ln
    \frac{Q^2}{\mu^2}}, \ \ \ \beta_2 = {11 \, N_c - 2 \, N_f\over 12
    \, N_c}.
\end{align}
In \eq{conf-proj} we are following the notation of Fadin and Lipatov
\cite{Fadin:1998py}: the function $\delta (\gamma)$ is rather involved
and is given by Eq.~(14) in \cite{Fadin:1998py} (for our definition of
rapidity, $Y = \ln (s/k k')$).

For future use we need to separate in \eq{conf-proj} the terms
symmetric and anti-symmetric under the $\gamma\leftrightarrow
1-\gamma$ interchange. While $\chi_0(\gamma)$ is explicitly symmetric
under such interchange, $\delta(\gamma)$ can be decomposed as
\cite{Fadin:1998py}
\begin{align}
  \label{eq:delta_def}
  \frac{\delta(\gamma)}{4} = -{1\over 2} \, \beta_2 \, \chi'_0(\gamma)
  + \chi_1(\gamma)
\end{align}
where $\chi'_0(\gamma)={d\over d\gamma}\chi_0(\gamma)$ generates the
anti-symmetric under $\gamma\leftrightarrow 1-\gamma$ term and
$\chi_1(\gamma)$ denotes the symmetric term and is given by
\cite{Fadin:1998py}
\begin{align}
  \label{eq:chi1}
  \chi_1(\gamma) = & - \beta_2 \, \frac{\chi_0^2 (\gamma)}{2} +
  \frac{5}{3} \, \beta_2 \, \chi_0 (\gamma) + \frac{1}{3} \, \left( 1
    - \frac{\pi^2}{4} \right) \, \chi_0 (\gamma) + \frac{3}{2} \,
  \zeta (3) \notag \\ & - \frac{\pi^2 \, \cos (\pi \, \gamma)}{4 \,
    \sin^2 (\pi \, \gamma) \, (1 - 2 \, \gamma)} \, \left[ 3 + \left(
      1 + \frac{N_f}{N_c^3} \right) \frac{2 + 3 \, \gamma - 3 \,
      \gamma^2}{3 + 4 \, \gamma - 4 \, \gamma^2} \right] - \frac{1}{4}
  \, \chi''_0 (\gamma) + \frac{\pi^3}{4 \, \sin (\pi \, \gamma)} -
  \phi (\gamma)
\end{align}
with $\zeta (z)$ the Riemann $\zeta$-function, $\chi''_0 (\gamma) =
{d^2\over d\gamma^2} \chi_0 (\gamma)$, and
\begin{align}
  \label{eq:phi}
  \phi (\gamma) = \sum_{n=0}^\infty \, (-1)^n \, \left[ \frac{\psi
      (n+1 +\gamma) - \psi (1)}{(n+\gamma)^2} + \frac{\psi (n+2 -
      \gamma) - \psi (1)}{(n+ 1 - \gamma)^2} \right].
\end{align}

Using the decomposition \eqref{eq:delta_def} we rewrite \eq{conf-proj}
as
\begin{align}
  \label{conf-proj2}
  \int d^2 q \, K^{{\rm LO}+ {\rm NLO}}(k,q) \ q^{2\gamma-2} = \left[
    \bar{\alpha}_\mu \, \chi_0 (\gamma) - \bar{\alpha}^2_\mu \,
    \beta_2 \, \chi_0(\gamma) \, \ln{k^2\over \mu^2} - \frac{1}{2} \,
    \bam^2 \, \beta_2 \, \chi'_0 (\gamma) + \bar{\alpha}_\mu^2 \,
    \chi_1 (\gamma) \right] \, k^{2\gamma-2}.
\end{align}

It is clear from \eq{conf-proj} (or from \eq{conf-proj2}) that the
eigenfunctions of the LO BFKL kernel (simple powers $k^{2\gamma-2}$
in the azimuthally symmetric case at hand) are not the eigenfunctions
of the NLO kernel, since the expression in the square brackets on the
right-hand-side of \eq{conf-proj} is also $k$-dependent. Notice also
that this property is entirely due to the running of the coupling
constant (and this is why the $k$-dependence in the square brackets on
the right of \eq{conf-proj} comes in with a factor of $\beta_2$): in
the conformal ${\cal N} =4$ SYM theory, where the coupling does not
run, eigenfunctions of the LO BFKL kernel are also eigenfunctions of
the arbitrary-order kernel, such that construction of the general form
of the solution is rather straightforward
\cite{Kotikov:2000pm,Balitsky:2008rc,Balitsky:2009xg,Balitsky:2009yp}.


\subsection{Constructing the Eigenfunctions}

As emphasized in the Introduction, we will find a set of
eigenfunctions for the LO+NLO BFKL kernel. Since $K^{{\rm LO}+ {\rm
    NLO}}$ is a perturbative expansion of the full BFKL equation
kernel, it appears logical to attempt constructing the LO+NLO BFKL
kernel's eigenfunctions perturbatively as well, order by order in the
coupling constant. The LO BFKL kernel eigenfunctions are known: these
are simply $k^{2\gamma-2}$. We will look for the LO+NLO BFKL kernel
eigenfunctions as perturbative corrections to this LO
eigenfunctions. We thus write the LO+NLO eigenfunctions as
\begin{align}
  \label{eq:eig1}
  H_\gamma(k) = k^{2\gamma-2} + \bar{\alpha}_\mu \, F_\gamma(k),
\end{align}
where the $F_\gamma(k)$ is a function to be determined in order to
make $H_\gamma(k)$ the eigenfunctions of the LO+NLO kernel.

Acting on $H_\gamma (k)$ with the LO+NLO kernel and employing
\eq{conf-proj2}, within the order-$\bam$ accuracy we have
\begin{align}
  \int & d^2q \, \Big[ K^{\rm LO}(k,q) + \bar{\alpha}_\mu \, K^{\rm
    NLO}(k,q)\Big] H_\gamma(q)
  \nonumber\\
  = & \, \chi_0(\gamma) \, k^{2\gamma-2} + \bar{\alpha}_\mu \int d^2 q
  \, K^{\rm LO}(k,q) F_\gamma(q) - \bar{\alpha}_\mu \, \beta_2 \,
  \chi_0(\gamma) \, k^{2\gamma-2}\ln{k^2\over \mu^2} +
  \bar{\alpha}_\mu \left( - \frac{1}{2} \, \beta_2 \, \chi'_0 (\gamma)
    + \chi_1 (\gamma) \right) \, H_\gamma(k)
  \nonumber\\
  = & \, \chi_0(\gamma) \, H_\gamma(k) - \bar{\alpha}_\mu \,
  \chi_0(\gamma) \, F_\gamma(k) + \bar{\alpha}_\mu \int d^2 q \,
  K^{\rm LO}(q,k) \, F_\gamma(q) - \bar{\alpha}_\mu \, \beta_2 \,
  \chi_0 (\gamma) \, k^{2\gamma-2} \ln {k^2\over \mu^2} \notag \\ & +
  \bar{\alpha}_\mu \left( - \frac{1}{2} \, \beta_2 \, \chi'_0 (\gamma)
    + \chi_1 (\gamma) \right) \, H_\gamma(k).
\label{condition}
\end{align}
From \eqref{condition} we deduce that $H_\gamma(k)$ is an
eigenfunction of the LO+NLO kernel if there exists a function
$c(\gamma)$ such that (up to order-$\bam$ corrections)
\begin{align}
  - \chi_0(\gamma) \, F_\gamma(k) + \int d^2 q \, K^{\rm LO}(k,q)
  F_\gamma(q) - \beta_2 \, \chi_0(\gamma) \, H_\gamma(k) \,
  \ln{k^2\over \mu^2} = c(\gamma) \, H_\gamma(k)
\label{condi}
\end{align}
so that we would have
\begin{align}\label{eig2}
  \int d^2 q \, K^{{\rm LO} + {\rm NLO}} (k,q) \, H_\gamma(q) = \left[
    \bar{\alpha}_\mu \, \chi_0(\gamma) + \bar{\alpha}_\mu^2 \left( -
      \frac{1}{2} \, \beta_2 \, \chi'_0 (\gamma) + \chi_1 (\gamma) +
      c(\gamma) \right) \right] H_\gamma(k) \equiv \Delta (\gamma) \,
  H_\gamma(k).
\end{align}
Here we have also defined a general notation, $\Delta (\gamma)$, for the
eigenvalue of the BFKL kernel.  Our task is to determine the functions
$c(\gamma)$ and $F_\gamma(k)$.

At the order we are working we can rewrite (\ref{condi}) as 
\begin{align}
  - \chi_0(\gamma) F_\gamma(k) + \int d^2 q \, K^{\rm LO}(k,q)
  F_\gamma(q) - \beta_2 \, \chi_0(\gamma) \, k^{2\gamma-2} \,
  \ln{k^2\over \mu^2} = c(\gamma) \, k^{2\gamma-2}
\label{condi1}
\end{align}
and deduce that $F_\gamma(k)$ has to be proportional to
$k^{2\gamma-2}$. Equation (\ref{condi1}) gives the condition that
$F_\gamma(k)$ has to satisfy in order for $H_\gamma(k)$ be an
eigenfunction of the NLO kernel. We look for a solution of
(\ref{condi1}) using the following ansatz
\begin{align}
  F_\gamma(k) = \sum_{n=0}^\infty c_n (\gamma) \left(\ln{k^2\over
      \mu^2}\right)^n \, k^{2\gamma-2},
\label{ansatz}
\end{align}
where $c(\gamma)$ are some (smooth) complex-valued functions of $\gamma$. 
Substituting \eq{ansatz} into \eq{condi1} and dividing everything by
$k^{2\gamma-2}$ yields
\begin{align}
  \label{ans1}
  - \chi_0(\gamma) \, \sum_{n=0}^\infty c_n (\gamma)
  \left(\ln{k^2\over \mu^2}\right)^n + \sum_{n=0}^\infty c_n (\gamma)
  \, k^{- 2 \gamma + 2} \, \int d^2 q \, K^{\rm LO}(k,q) \,
  \left(\ln{q^2\over \mu^2}\right)^n \, q^{2\gamma-2} = c(\gamma) +
  \beta_2 \, \chi_0(\gamma) \, \ln{k^2\over \mu^2}.
\end{align}
Rewriting the logarithms as derivatives with respect to $\gamma$ in
the second term on the left of \eq{ans1} we get
\begin{align}
  \label{ans2}
  \sum_{n=0}^\infty c_n (\gamma) \, \sum_{m=0}^{n-1} \left(
    \stackeven{n}{m} \right) \, \chi_0^{(n-m)} (\gamma) \, \ln^m
  {k^2\over \mu^2} = c(\gamma) + \beta_2 \, \chi_0(\gamma) \,
  \ln{k^2\over \mu^2}
\end{align}
where $\chi_0^{(n)} (\gamma) = d^{n} \chi_0 (\gamma) / d \gamma^{n}$.

To find the solution of \eqref{ans2} we truncate the series
\eqref{ansatz} at $n=2$, that is we put $c_3 (\gamma) = c_4 (\gamma) =
c_5 (\gamma) = \ldots = 0$, leaving only $c_0 (\gamma)$, $c_1
(\gamma)$, and $c_2 (\gamma)$ non-zero. As we will shortly see, this
truncation allows us to find a solution for $F_\gamma (k)$, that is a
set of eigenfunctions for the LO+NLO BFKL kernel. Note that our goal
is to find a complete ortho-normal set of eigenfunctions of
$K^{LO+NLO}$. Therefore it does not matter whether our technique of
finding the solution of \eqref{ans2} is general enough as long as a
complete set of eigenfunctions is found.

Keeping only $c_0 (\gamma)$, $c_1 (\gamma)$, and $c_2 (\gamma)$
non-zero in \eq{ans2} we obtain
\begin{align}
  c_1(\gamma) \, \chi_0' (\gamma) + c_2(\gamma) \left[ \chi_0''
    (\gamma) + 2 \, \chi_0 '(\gamma) \, \ln{k^2\over \mu^2} \right] =
  c(\gamma) + \beta_2 \, \chi_0(\gamma) \, \ln{k^2\over \mu^2}.
\label{condi2}
\end{align} 
Since \eq{condi2} has to be valid at all values of $k$, we equate the
coefficients of the terms proportional to $\ln{k^2\over \mu^2}$ on
both sides of \eqref{condi2} to get
\begin{align}\label{c2}
  c_2(\gamma) = {\beta_2 \, \chi_0 (\gamma) \over 2 \, \chi_0
    '(\gamma)}.
\end{align}
The remaining terms in \eqref{condi2} (without $\ln{k^2\over \mu^2}$)
give
\begin{align}
  c(\gamma) = c_1(\gamma) \, \chi_0'(\gamma) + {\beta_2 \, \chi_0
    (\gamma) \, \chi_0 ''(\gamma) \over 2 \, \chi_0'(\gamma)}.
\label{cg}
\end{align}
Substituting $c_2$ from \eqref{c2} along with $c_3 = c_4 = \ldots = 0$
into Eqs.~\eqref{ansatz} and \eqref{eq:eig1} we obtain the
eigenfunctions of the LO+NLO BFKL kernel,
\begin{align}
  H_\gamma(k) = k^{2\gamma-2} \, \left[ 1 + \bar{\alpha}_\mu \left(
      {\beta_2 \, \chi_0 (\gamma) \over 2 \, \chi'_0 (\gamma)} \,
      \ln^2{k^2\over \mu^2} + c_1(\gamma) \ln{k^2\over \mu^2} +
      c_0(\gamma) \right) \right],
\label{eigenf}
\end{align}
for any regular and smooth complex-valued functions $c_0(\gamma)$ and
$c_1(\gamma)$. The corresponding eigenvalues $\Delta (\gamma)$ can be
read off from Eqs.~\eqref{eig2} and \eqref{cg}.

It is worth noticing that although we searched for a correction to the
eigenfunction in $\bam$, we ended up automatically with a correction
proportional to $\bam \, \beta_2$. This is indicative of the fact that
we are actually expanding around the conformal point. This is not
surprising, since the terms breaking conformal invariance and
therefore making the LO BFKL kernel eigenfunctions not eigenfunctions
of the NLO BFKL kernel, are proportional to $\beta_2$. Hence it is
natural to expect that the LO eigenfunction gets corrections at NLO
proportional to $\beta_2$.

Since $c_0(\gamma)$ and $c_1(\gamma)$ are arbitrary smooth functions
of $\gamma$ we have a set of eigenfunctions in \eq{eigenf} for each
$c_0(\gamma)$ and $c_1(\gamma)$. By requiring that the eigenfunctions
$H_\gamma(k)$ satisfy the completeness and orthogonality relations at
order $\alpha_\mu$ would reduce the freedom in these functions by
generating constraints on $c_1 (\gamma)$ and/or $c_0 (\gamma)$.


\subsection{Completeness and Orthogonality}

The completeness relation for the $H_\gamma(k)$ eigenfunctions is
\begin{align}
  \int\limits_{\sigma-i\infty}^{\sigma+i\infty} {d\gamma\over 2\pi i}
  \, H_\gamma(k) \, H^*_\gamma(k') = \delta(k^2-k'^2)
\label{NLO-compl}
\end{align}
where $\sigma$ is a real parameter that has to be determined so that
\eq{NLO-compl} is satisfied and the asterisk denotes complex
conjugation. We need to substitute $H_\gamma(k)$ from \eq{eigenf} into
\eq{NLO-compl} and make sure the latter is satisfied order-by-order in
$\bam$.

The leading-order term in \eq{eigenf} already satisfies \eq{NLO-compl}
for
\begin{align}
  \label{gamma_nu}
  \gamma={1\over 2} + i \, \nu
\end{align}
with $\nu$ a real variable. This fixes $\sigma={1\over 2}$, such that
\eq{NLO-compl} becomes
\begin{align}
  \label{NLO-compl3}
  \int\limits_{\frac{1}{2}-i\infty}^{\frac{1}{2}+i\infty}
  {d\gamma\over 2\pi i} \, H_\gamma(k) \, H^*_\gamma(k') =
  \delta(k^2-k'^2).
\end{align}

Note that at LO the power-law functions $k^{2 \gamma - 2}$ are
eigenfunctions of the LO BFKL kernel for any $\gamma$ (with $0 < {\rm
  Re} \, \gamma <1$). However for $\gamma={1\over 2}+i \, \nu$ with
$\nu$ a real parameter, these power-law functions form an ortho-normal
set in the functional space (of LO BFKL solutions) upon which the
action of the LO kernel is well-defined and yields normalized
functions. (That is, the integrals like
\begin{align}
  \label{int_def}
  \int d^2 k \, d^2 q \, f^* (k) \, K(k,q) \, g(q)
\end{align}
are finite for any functions $f(k)$, $g(k)$ from the set.)
Consequently, the LO BFKL kernel acts on powers $k^{-1+2 \, i \, \nu}$
as a hermitean operator with real eigenvalues. Hermiticity of the BFKL
kernel has a straightforward physical interpretation: BFKL ladder
(drawn vertically) is up-down symmetric. Formally this means that the
BFKL kernel satisfies the following property $K (k,q) = K(q,k)$ both
at LO and at NLO
\cite{Kuraev:1977fs,Kuraev:1976ge,Balitsky:1978ic,Fadin:1998py,Ciafaloni:1998gs},
which, when combined with the fact that the kernel is real, implies
that it is hermitean. One therefore expects that the eigenvalues of
the BFKL kernel for $\gamma={1\over 2}+i \, \nu$ should be real to any
order in the coupling constant. For the LO eigenvalues
\eqref{eq:LOeig} taken at $\gamma={1\over 2}+i \, \nu$ this property
is obviously satisfied:
\begin{align}
  \label{chi_0nu}
  \chi_0 (\nu) \equiv \chi_0 \left( \gamma={1\over 2} + i \, \nu
  \right) = 2 \, \psi(1) - \psi \left({1\over 2} + i \, \nu \right) -
  \psi \left({1\over 2} - i \, \nu \right)
\end{align}
is manifestly real for real $\nu$. We will shortly see that the
reality condition is satisfied by the NLO eigenvalues from \eq{eig2}.

In the following, it will be more convenient to keep track of the real
and imaginary parts of the expressions by switching to the $\nu$
variable using \eq{gamma_nu}. To that end we rewrite \eq{eigenf} as
\begin{align}
  H_{\frac{1}{2}+i \, \nu} (k) = k^{-1 + 2\, i \, \nu} \, \left[ 1 +
    \bar{\alpha}_\mu \left( i \, {\beta_2 \, \chi_0 (\nu) \over 2 \,
        \chi'_0 (\nu)} \, \ln^2{k^2\over \mu^2} + c_1(\nu)
      \ln{k^2\over \mu^2} + c_0(\nu) \right) \right],
\label{eigenf_nu}
\end{align}
where we defined $c_j (\nu) \equiv c_j(\gamma(\nu)) = {\rm Re} \,
[c_j(\nu)] + i \, {\rm Im} \, [c_j(\nu)]$ for $j = 0, 1$ and $\chi'_0
(\nu) = d \chi_0 (\nu)/d\nu$.

Substituting \eqref{eigenf_nu} into \eqref{NLO-compl3} we obtain the
following condition at order-$\bam$:
\begin{align}
  \int\limits_{-\infty}^\infty d\nu \left({k^2\over
      k'^2}\right)^{i\nu} \left[ i \, {\beta_2 \, \chi_0 (\nu) \over
      \chi_0'(\nu)} \, \ln{kk'\over \mu^2} \, \ln{k^2\over k'^2} + 2
    \, {\rm Re}[c_1(\nu)] \, \ln{kk'\over \mu^2} + i \, {\rm
      Im}[c_1(\nu)] \, \ln{k^2\over k'^2} + 2 \, {\rm Re}[c_0(\nu)]
  \right] = 0.
\label{NLO-compl1}
\end{align}
The first thing to notice in \eq{NLO-compl1} is that ${\rm
  Im}[c_0(\nu)]$ does not get constrained by the completeness relation
since it cancels out in the $H_{\frac{1}{2}+i \, \nu} (k)
[H_{\frac{1}{2}+i \, \nu} (k')]^*$ product at order-$\bam$.  From
(\ref{NLO-compl1}) it is also clear that the terms that are
$\mu$-dependent should vanish independently from the
rest:\footnote{Since $c_1 (\nu)$ and $c_0 (\nu)$ are dimensionless and
  momentum-independent, they can not depend on $\mu$.} this gives
\begin{align}
  0 = \int\limits_{-\infty}^\infty d\nu \left({k^2\over
      k'^2}\right)^{i\nu} \left[i \, {\beta_2 \, \chi_0 (\nu) \over
      \chi_0'(\nu)} \, \ln{k^2\over k'^2} + 2 \, {\rm Re}[c_1(\nu)]
  \right] = \int\limits_{-\infty}^\infty d\nu \left({k^2\over
      k'^2}\right)^{i\nu} \left[ -
    {\partial\over\partial\nu}\left({\beta_2 \, \chi_0(\nu)\over
        \chi_0'(\nu)}\right) + 2 \, {\rm Re}[c_1(\nu)] \right].
\label{cond1}
\end{align}
\eq{cond1} is satisfied for all $k$ and $k'$ only if
\begin{align} {\rm Re}[c_1(\nu)] = \frac{\beta_2}{2} \,
  {\partial\over\partial\nu} \left( {\chi_0 (\nu) \over \chi_0'(\nu)}
  \right) = \frac{\beta_2}{2} \left( 1 - \frac{\chi_0 (\nu) \,
      \chi_0'' (\nu)}{\chi_0' (\nu)^2} \right).
\label{condition1}
\end{align}
We have determined ${\rm Re}[c_1(\nu)]$. Notice that the integral in
(\ref{cond1}) contains a divergence at $\nu=0$ due to $\chi_0' (\nu)$
in the denominator of the first term: we assume that the divergence is
regularized using the principle value prescription.  Now, using
(\ref{condition1}) in (\ref{NLO-compl1}) we have
\begin{align}
  0 = \int\limits_{-\infty}^\infty d\nu \left({k^2\over
      k'^2}\right)^{i\nu} \left[ i \, {\rm Im}[c_1(\nu)] \,
    \ln{k^2\over k'^2} + 2 \, {\rm Re}[c_0(\nu)] \right] =
  \int\limits_{-\infty}^\infty d\nu \left({k^2\over
      k'^2}\right)^{i\nu} \left[ - {\partial\over \partial
      \nu}\left({\rm Im}[c_1(\nu)]\right) + 2 \, {\rm Re}[c_0(\nu)]
  \right]
\label{NLO-compl2}
\end{align}
which is satisfied for all $k$ and $k'$ only if
\begin{align} 
  - {\partial\over \partial \nu}\left( {\rm Im}[c_1(\nu)] \right) + 2
  \, {\rm Re}[c_0(\nu)]=0.
\label{cond2}
\end{align}
We see that the completeness relation \eqref{NLO-compl} is not
sufficient to completely fix $c_0(\nu)$ (with ${\rm Im} [c_1(\nu)]$
fixed by ${\rm Re} [c_0(\nu)]$ through \eqref{cond2}).

Keeping this in mind, and using Eqs.~\eqref{condition1} and
\eqref{cond2} in \eq{eigenf_nu} we arrive at the following form of the
LO+NLO BFKL kernel eigenfunctions
\begin{align}
  H_{\frac{1}{2}+i \, \nu} (k) = k^{-1 + 2\, i \, \nu} \, \left[ 1 +
    \bar{\alpha}_\mu \left( i \, {\beta_2 \, \chi_0 (\nu) \over 2 \,
        \chi'_0 (\nu)} \, \ln^2{k^2\over \mu^2} + \frac{\beta_2}{2}
      \left( 1 - \frac{\chi_0 (\nu) \, \chi_0'' (\nu)}{\chi_0'
          (\nu)^2} \right) \ln{k^2\over \mu^2} + i \, {\rm
        Im}[c_1(\nu)] \ln{k^2\over \mu^2} + c_0(\nu) \right)
  \right].
\label{eigenf_nu2}
\end{align}
The eigenvalues $\Delta (\nu) \equiv \Delta (\gamma(\nu))$ of the
eigenfunctions \eqref{eigenf_nu2} can be easily constructed using
Eqs.~\eqref{eig2} and \eqref{cg}:
\begin{align}
  \label{eigenv1}
  \Delta (\nu) & = \bar{\alpha}_\mu \, \chi_0(\nu) +
  \bar{\alpha}_\mu^2 \left( \frac{i}{2} \, \beta_2 \, \chi'_0 (\nu) +
    \chi_1 (\nu) + c(\nu) \right)\notag \\ & = \bar{\alpha}_\mu \,
  \chi_0(\nu) + \bar{\alpha}_\mu^2 \left( \frac{i}{2} \, \beta_2 \,
    \chi'_0 (\nu) + \chi_1 (\nu) - i \, c_1 (\nu) \, \chi_0' (\nu) - i
    \, \frac{\beta_2 \, \chi_0 (\nu) \, \chi''_0 (\nu)}{2 \, \chi'_0
      (\nu)} \right).
\end{align}
Here $\chi_1 (\nu) \equiv \chi_1 (\gamma (\nu))$ and $c (\nu) \equiv c
(\gamma (\nu))$.  Finally, with the help of \eqref{condition1} we
rewrite \eq{eigenv1} as
\begin{align}
  \label{eigenv2}
  \Delta (\nu) = \bar{\alpha}_\mu \, \chi_0(\nu) + \bar{\alpha}_\mu^2
  \, \bigg[ \chi_1 (\nu) + {\rm Im}[c_1 (\nu)] \, \chi_0' (\nu)
  \bigg].
\end{align}
We see that the LO+NLO eigenvalues are manifestly real, as expected
from eigenvalues of a hermitean operator such as the LO+NLO BFKL
kernel! Note also that the reality of the eigenvalues was achieved
after the cancellation of the $(\gamma \leftrightarrow 1-\gamma)$-odd
term from \eq{eq:delta_def} calculated in \cite{Fadin:1998py}: as
$\gamma \leftrightarrow 1-\gamma$ corresponds to $\nu \leftrightarrow
- \nu$, this term is in fact imaginary, and it had to cancel for us to
obtain a real eigenvalue.

Eqs.~\eqref{eigenf_nu2} and \eqref{eigenv2} give us the LO+NLO BFKL
kernel eigenfunctions and their corresponding eigenvalues up to a
freedom of choosing $c_0 (\nu)$ (with ${\rm Im}[c_1 (\nu)]$ fixed by
\eqref{cond2}). Below we will show that the solution of LO+NLO BFKL
equation does not depend on $c_0 (\nu)$ (and ${\rm Im}[c_1 (\nu)]$),
such that one is always free to put them to zero (both in the
eigenfunctions \eqref{eigenf_nu2} and in the eigenvalues
\eqref{eigenv2}) and no ambiguity is left.

In Appendix A we show that the eigenfunctions \eqref{eigenf_nu2} also
satisfy the orthogonality condition
\begin{align}
  \label{NLO-ortho1}
  \int d^2 k \, H_{\frac{1}{2} + i \, \nu}(k) \, \left[ H_{\frac{1}{2}
      + i \, \nu'} (k) \right]^* = 2 \, \pi^2 \, \delta(\nu - \nu').
\end{align}
Note that orthogonality does not impose any additional constraints on
$c_0 (\nu)$, thus leaving it unspecified.


\subsection{The Phase and $\nu$-Reparametrization Freedom} 

Before we clarify the origin of the remaining freedom of choosing $c_0
(\nu)$ in the eigenfunctions \eqref{eigenf_nu2}, let us note that we
have found eigenfunctions of the LO+NLO BFKL kernel satisfying the
completeness \eqref{NLO-compl3} and orthogonality \eqref{NLO-ortho1}
conditions: however, these conditions do not yet fix the
eigenfunctions uniquely. In this Subsection we will show that there
are two trivial degrees of freedom in the choice of eigenfunctions,
one due to a choice of the overall phase, while another one is due to
a choice of the variable $\nu$. These symmetries of the problem will
allow us to always choose LO+NLO BFKL eigenfunctions with $c_0 (\nu)
=0$ and ${\rm Im} [c_1 (\nu)] =0$.

An eigenfunction of BFKL kernel remains an eigenfunction after a
rescaling by a constant. \eq{NLO-ortho1}, being an ortho-normality
condition, limits the rescaling freedom for functions $H_{\frac{1}{2}
  + i \, \nu}(k)$ to
\begin{align}
  \label{resc1}
  H_{\frac{1}{2} + i \, \nu}(k) \, \to \, e^{i \, \theta} \,
  H_{\frac{1}{2} + i \, \nu}(k)
\end{align}
with any real phase $\theta$. Note that the same phase freedom exists
already in the LO BFKL eigenfunctions, since we can always rescale
$k^{-1 + 2 \, i \, \nu} \to e^{i \, \theta} \, k^{-1 + 2 \, i \,
  \nu}$. The phase $\theta$ is fixed by convention: at the LO it is usually
put to zero. Similarly at NLO we can always remove the phase of
$H_{\frac{1}{2} + i \, \nu}(k)$ by rescaling
\begin{align}
  \label{resc2}
  H_{\frac{1}{2} + i \, \nu}(k) \, \to \, e^{- i \, \bam \, \left(
      {\rm Im}[c_0(\nu)] - {\rm Im}[c_1 (\nu)] \, \ln \mu^2 \right) }
  \, H_{\frac{1}{2} + i \, \nu}(k) \approx \left[ 1 - i \, \bam \, \,
    \left( {\rm Im}[c_0(\nu)] - {\rm Im}[c_1 (\nu)] \, \ln \mu^2
    \right) \right] \, H_{\frac{1}{2} + i \, \nu}(k).
\end{align}
That is we can always choose the phase convention in which ${\rm
  Im}[c_0(\nu)]$ is zero and the LO+NLO BFKL eigenfunctions are
\begin{align}
  H_{\frac{1}{2}+i \, \nu} (k) = k^{-1 + 2\, i \, \nu} \, \left[ 1 +
    \bar{\alpha}_\mu \left( i \, {\beta_2 \, \chi_0 (\nu) \over 2 \,
        \chi'_0 (\nu)} \, \ln^2{k^2\over \mu^2} + \frac{\beta_2}{2}
      \left( 1 - \frac{\chi_0 (\nu) \, \chi_0'' (\nu)}{\chi_0'
          (\nu)^2} \right) \ln{k^2\over \mu^2} + i \, {\rm
        Im}[c_1(\nu)] \ln{k^2} + {\rm Re}[c_0 (\nu)] \right) \right]
\label{eigenf_nu3}
\end{align}
with the eigenvalues still given by \eq{eigenv2}. \footnote{If one is
  uncomfortable with logarithms of dimensionful quantities, such as
  $\ln{\mu^2}$ in \eqref{resc2} and $\ln{k^2}$ in \eqref{eigenf_nu3},
  the arguments in this Section can be repeated for dimensionless
  eigenfunctions defined by ${\tilde H}_{\frac{1}{2}+i \, \nu} (k)
  \equiv \mu^{1 - 2 \, i \, \nu} H_{\frac{1}{2}+i \, \nu} (k)$ if one
  removes ${\rm Im}[c_1 (\nu)]$ from \eq{resc2} while leaving
  \eq{nu_shift} the same.}

Let us show now how to eliminate ${\rm Im}[c_1(\nu)]$ and ${\rm
  Re}[c_0 (\nu)]$ in \eq{eigenf_nu3} by using another degree of
freedom, namely the choice of reparametrization for the variable
$\nu$ labeling the eigenfunctions. For instance, let us
shift the $\nu$-variable by defining
\begin{align}
  \label{nu_shift}
 \nu' = \nu + \bam \, {\rm Im}[c_1(\nu)]
\end{align}
in the completeness relation \eqref{NLO-compl3} with the $H$-function
given in (\ref{eigenf_nu3}). We obtain (to order-$\bam$)
\begin{align}
  \delta(k^2-k'^2)&=\int\limits^{+\infty}_{-\infty}\,{d\nu\over 2\pi}
  \,
  H_{\frac{1}{2} + i \, \nu}(k)\,[H_{\frac{1}{2} + i \, \nu}(k')]^*\label{comp-nu1} \\
  &= \int\limits^{+\infty}_{-\infty} \,{d\nu'\over 2\pi} \,{d\nu\over
    d\nu'}\,{1\over k\,k'} \left({k^ 2\over k'^2}\right)^{i \, \nu'}\,
  \left(1-i \, \bam \,{\rm Im}[c_1(\nu')]\ln k^2\right)
  \left(1+i \, \bam\,{\rm Im}[c_1(\nu')] \ln k'^2 \right)\notag\\
  & \times \left[ 1 + \bar{\alpha}_\mu \left( i \, {\beta_2 \, \chi_0
        (\nu') \over 2 \, \chi'_0 (\nu')} \, \ln^2{k^2\over \mu^2} +
      \frac{\beta_2}{2} \left( 1 - \frac{\chi_0 (\nu') \, \chi_0''
          (\nu')}{\chi_0'(\nu')^2} \right) \ln{k^2\over \mu^2}
      + i \, {\rm Im}[c_1(\nu')] \ln{k^2} + {\rm Re}[c_0 (\nu')] \right) \right] \notag \\
  & \times \left[ 1 + \bar{\alpha}_\mu \left( -i \, {\beta_2 \, \chi_0
        (\nu') \over 2 \, \chi'_0 (\nu')} \, \ln^2{k'^2\over \mu^2} +
      \frac{\beta_2}{2} \left( 1 - \frac{\chi_0 (\nu') \, \chi_0''
          (\nu')}{\chi_0' (\nu')^2} \right) \ln{k'^2\over \mu^2} -i \,
      {\rm Im}[c_1(\nu')] \ln{k'^2} + {\rm Re}[c_0 (\nu')] \right)
  \right]. \notag
\end{align}
Using 
\begin{align}
  \label{jacobian}
  \frac{d \nu}{d \nu'} = 1 - \bam \, \frac{\pd}{\pd \nu'} \, {\rm
    Im}[c_1(\nu')]
\end{align}
and the relation (\ref{cond2}) we recast Eq.~(\ref{comp-nu1}) as
\begin{align}
  \delta(k^2-k'^2)=\int\limits^{+\infty}_{-\infty} {d\nu'\over 2\pi}
  \, {1\over k\,k'} \, \left({k^2\over k'^2}\right)^{i \, \nu'} & \left[
    1 + \bar{\alpha}_\mu \left( i \, {\beta_2 \, \chi_0 (\nu') \over 2
        \, \chi'_0 (\nu')} \, \ln^2{k^2\over \mu^2} +
      \frac{\beta_2}{2} \left( 1 - \frac{\chi_0 (\nu') \, \chi_0''
          (\nu')}{\chi_0'
          (\nu')^2} \right) \ln{k^2\over \mu^2} \right) \right]\notag\\
  \times & \left[ 1 + \bar{\alpha}_\mu \left( -i \, {\beta_2 \, \chi_0
        (\nu') \over 2 \, \chi'_0 (\nu')} \, \ln^2{k'^2\over \mu^2} +
      \frac{\beta_2}{2} \left( 1 - \frac{\chi_0 (\nu') \, \chi_0''
          (\nu')}{\chi_0' (\nu')^2} \right) \ln{k'^2\over \mu^2}
    \right) \right].\label{comp-nu2}
\end{align}
Therefore, we may define the new LO+NLO BFKL eigenfunctions after the
$\nu$-reparametrization \eqref{nu_shift} as
\begin{align} 
H_{\frac{1}{2}+i \, \nu} (k) = k^{-1 + 2\, i \, \nu} \, \left[ 1 +
    \bar{\alpha}_\mu \, \beta_2 \left( i \, {\chi_0 (\nu) \over 2 \,
        \chi'_0 (\nu)} \, \ln^2{k^2\over \mu^2} + \frac{1}{2} \left( 1
        - \frac{\chi_0 (\nu) \, \chi_0'' (\nu)}{\chi_0' (\nu)^2}
      \right) \ln{k^2\over \mu^2} \right) \right].
          \label{eigenf_nu_final}
\end{align}
One can see from \eq{comp-nu2} that these new functions satisfy the
completeness relation \eqref{NLO-compl3}.  The functions
\eqref{eigenf_nu_final} may be obtained from \eq{eigenf_nu3} by
putting ${\rm Im}[c_1(\nu)] = 0$ and ${\rm Re}[c_0 (\nu)] =0$ in
it. This shows that functions $H_{\frac{1}{2}+i \, \nu}$ in
\eqref{eigenf_nu_final} are the eigenfunctions of the LO+NLO BFKL
kernel. We have eliminated ${\rm Im}[c_1(\nu)]$ and ${\rm Re}[c_0
(\nu)]$ in \eqref{eigenf_nu3} by simply using the freedom
\eqref{nu_shift} to redefine $\nu$. One can easily see that the
functions \eqref{eigenf_nu_final} also satisfy the orthogonality
relation \eqref{NLO-ortho1}.

It is also important to note that the $\nu = \nu' - \bam \, {\rm
  Im}[c_1(\nu')]$ transformation eliminates ${\rm Im}[c_1(\nu)]$ from
the eigenvalue \eqref{eigenv2} since, to order-$\bam$,
\begin{align}
  \label{eigenv_rep}
  \Delta (\nu) = \bar{\alpha}_\mu \, \chi_0(\nu) + \bar{\alpha}_\mu^2
  \, \bigg[ \chi_1 (\nu) + {\rm Im}[c_1 (\nu)] \, \chi_0' (\nu) \bigg]
  = \bar{\alpha}_\mu \, \chi_0(\nu') + \bar{\alpha}_\mu^2 \, \chi_1
  (\nu') + {\cal O} (\bam^3).
\end{align}
We see that ${\rm Im}[c_1(\nu)]$ in the eigenvalue \eqref{eigenv2} is
closely related to the choice of the $\nu$-variable, and can be
eliminated by such choice.

We can always utilize the $\nu$-reparametrization invariance to
remove ${\rm Im}[c_1(\nu)]$ and ${\rm Re}[c_0 (\nu)]$ in
Eqs.~\eqref{eigenf_nu3} and \eqref{eigenv2}: note that this procedure
leaves the solution unchanged. Our final form of the LO+NLO BFKL
eigenfunctions $H_{{1\over 2}+i\nu}$ is given in \eq{eigenf_nu_final}.
The corresponding eigenvalues are obtained from \eqref{eigenv2} by
using ${\rm Im}[c_1(\nu)] =0$, such that
\begin{align}
  \label{eigenv3}
  \Delta (\nu) = \bar{\alpha}_\mu \, \chi_0(\nu) + \bar{\alpha}_\mu^2
  \, \chi_1 (\nu).
\end{align}


\subsection{NLO BFKL Solution}

Using the eigenfunctions \eqref{eigenf_nu_final} and eigenvalues
\eqref{eigenv3} of the LO+NLO BFKL kernel, along with the completeness
relation \eqref{NLO-compl3} it is straightforward to write the solution
of \eq{eq:BFKL} with the initial condition \eqref{eq:init} to the NLO
accuracy as
\begin{align}
  \label{NLOsol}
  G (k, k', Y) = \int\limits_{-\infty}^{\infty} {d\nu \over 2\pi^2} \,
  e^{\left[\bar{\alpha}_\mu \, \chi_0(\nu) + \bar{\alpha}_\mu^2 \,
      \chi_1 (\nu)\right] \, Y} \, H_{\frac{1}{2}+i \, \nu} (k) \,
  \left[ H_{\frac{1}{2}+i \, \nu} (k') \right]^*.
\end{align}
Here $\chi_0(\nu)$ is given by \eq{chi_0nu} and, for completeness, we
rewrite \eq{eq:chi1} in terms of the $\nu$ variable as
\begin{align}
  \label{chi_1nu}
  \chi_1(\nu) = & - \beta_2 \, \frac{\chi_0^2 (\nu)}{2} + \frac{5}{3}
  \, \beta_2 \, \chi_0 (\nu) + \frac{1}{3} \, \left( 1 -
    \frac{\pi^2}{4} \right) \, \chi_0 (\nu) + \frac{3}{2} \, \zeta (3)
  \notag \\ & - \frac{\pi^2 \, \sinh (\pi \, \nu)}{8 \, \nu \, \cosh^2
    (\pi \, \nu) } \, \left[ 3 + \left( 1 + \frac{N_f}{N_c^3} \right)
    \frac{11 + 12 \, \nu^2}{16 \, (1+\nu^2)} \right] + \frac{1}{4} \,
  \chi''_0 (\nu) + \frac{\pi^3}{4 \, \cosh (\pi \, \nu)} - \phi
  \left(\frac{1}{2} + i \, \nu \right)
\end{align}
with $\phi$ given by \eq{eq:phi}. The eigenfunctions $H_{\frac{1}{2}+i
  \, \nu} (k)$ are given by \eq{eigenf_nu_final}.

\eq{NLOsol}, along with Eqs.~\eqref{eigenf_nu_final} and
\eqref{eigenv3}, are the main results of this work.


\section{General Form of the Solution of Higher-Order BFKL equation}

\label{sec:genform}

Above we have devised a solution of the NLO BFKL equation by obtaining
the LO+NLO BFKL kernel eigenfunctions using a perturbative expansion
around the LO conformal eigenfunctions. The resulting expression
\eqref{NLOsol} contains two perturbative expansions: one in the
eigenfunctions and another one in the exponent (the eigenvalue). While
this is different from a solution for the DGLAP evolution equation
\cite{Dokshitzer:1977sg,Gribov:1972ri,Altarelli:1977zs}, which only
has perturbative expansion in the anomalous dimensions (that is, in
the exponent giving the power of $Q^2$), the expansions utilized in
arriving at \eqref{NLOsol} are also well-defined and are under
theoretical control.

Based on the success of our strategy at the NLO level, we conjecture
that it can be applied to the generalized BFKL equation with the
kernel calculated to an arbitrary high order in the coupling
constant. The solution of the all-order BFKL equation can be formally
written as
\begin{align}
  \label{all_sol}
  G (k, k', Y) = \int\limits_{-\infty}^{\infty} {d\nu \over 2\pi^2} \,
  e^{\left[\bar{\alpha}_\mu \, \chi_0(\nu) + \bar{\alpha}_\mu^2 \,
      \chi_1 (\nu) + \bar{\alpha}_\mu^3 \,
      \chi_2 (\nu) + \ldots \right] \, Y} \, H_{\frac{1}{2}+i \, \nu} (k) \,
  \left[ H_{\frac{1}{2}+i \, \nu} (k') \right]^*
\end{align}
with the eigenfunctions
\begin{align}
  \label{eig_all}
  H_{\frac{1}{2}+i \, \nu} (k) = k^{-1 + 2\, i \, \nu} \, \left[ 1 +
    \bar{\alpha}_\mu \, \beta_2 \left( i \, {\chi_0 (\nu) \over 2 \,
        \chi'_0 (\nu)} \, \ln^2{k^2\over \mu^2} + \frac{1}{2} \left( 1
        - \frac{\chi_0 (\nu) \, \chi_0'' (\nu)}{\chi_0' (\nu)^2}
      \right) \ln{k^2\over \mu^2} \right) + \bam^2 \, f_2 \left(
      \frac{k}{\mu}, \nu \right) + \ldots \right].
\end{align}
Here $\chi_2 (\nu)$ and higher-order coefficients indicated by the
ellipsis in the exponent are the scale-invariant (conformal) $(\nu
\leftrightarrow -\nu)$-even (real-valued) parts of the prefactor
function generated by the action of the next-to-next-to-leading-order
(NNLO) (and higher-order) kernels on the LO eigenfunctions
(cf. \eq{conf-proj2}). The function $f_2 (k/\mu, \nu)$ denotes the
NNLO corrections to the eigenfunctions, with the ellipsis in the
expression \eqref{eig_all} for the eigenfunction denoting higher-order
corrections. Knowledge of the LO and NLO BFKL kernels is sufficient to
construct $f_2 (k/\mu, \nu)$, but is left for future work
\cite{GiovanniYuri2}.


\section{Properties of the NLO Solution}
\label{sec:NNLOans}

\subsection{On $\mu$-independence of the NLO solution}

First let us note that the BFKL kernel is independent of the arbitrary
renormalization scale $\mu$ at the order within the precision of the
perturbative calculation \cite{Fadin:1998py,Ciafaloni:1998gs}. Since
the $H_\gamma(k)$ eigenfunctions diagonalize the LO+NLO BFKL kernel,
we may write
\begin{align}
  \bam \, K^{\rm LO}(k,q) + \bam^2 \, K^{\rm NLO}(k, q) =
  \int\limits^{\infty}_{-\infty} {d\nu \over 2 \, \pi^2} \,
  \Delta(\nu) \, H_{\frac{1}{2} + i \, \nu} (k) \left[ H_{\frac{1}{2}
      + i \, \nu} (q) \right]^*.
\label{NLO-diag}
\end{align}
From (\ref{NLO-diag}) it is easy to see that since the kernel on the
left is $\mu$-independent, so should be the right-hand side of the
expression. Iterating the action of the kernel in \eq{NLO-diag} many
times we see that (in a somewhat schematic notation)
\begin{align}
  \label{NLO-diag2}
  \left[ \bam \, K^{\rm LO} + \bam^2 \, K^{\rm NLO} \right]^n (k, q) =
  \int\limits^{\infty}_{-\infty} {d\nu \over 2 \, \pi^2} \, \Delta^n
  (\nu) \, H_{\frac{1}{2} + i \, \nu} (k) \left[ H_{\frac{1}{2} + i \,
      \nu} (q) \right]^*
\end{align}
for any positive integer $n$.  Taking \eq{NLO-compl3} into account we
see that the integral on the right of \eqref{NLO-diag2} is
$\mu$-independent (within the precision of the approximation, that is,
up to and including order-$\bam^{n+1}$ terms) for any integer power $n
\ge 0$. Expanding \eq{NLOsol} in the powers of $Y$ we see that the
coefficients of this series are proportional to the right-hand-side of
\eq{NLO-diag2} with $n \ge 0$, and are, therefore, $\mu$-independent.

The $\mu$-independence can be seen even more explicitly by
substituting the eigenfunctions \eqref{eigenf_nu_final} into
\eq{NLOsol}. This yields
\begin{align}
  \label{NLOsol2}
  G (k, k', Y) = \int\limits_{-\infty}^{\infty} {d\nu \over 2\pi^2 \,
    k \, k'} \, e^{\left[\bar{\alpha}_\mu \, \chi_0(\nu) +
      \bar{\alpha}_\mu^2 \, \chi_1 (\nu)\right] \, Y} \, \left(
    \frac{k^2}{k'^2} \right)^{i \, \nu} \left[ 1 + \bar{\alpha}_\mu \,
    \beta_2 \left( i \, {\chi_0 (\nu) \over \chi'_0 (\nu)} \, \ln{k \,
        k' \over \mu^2} \, \ln {k^2 \over k'^2} + \frac{\pd}{\pd \nu}
      \left( \frac{\chi_0 (\nu) }{\chi_0' (\nu)} \right) \ln{k \, k'
        \over \mu^2} \right) \right],
\end{align}
to order-$\bam$ in the product of the eigenfunctions. In the first
term in the parenthesis of \eqref{NLOsol2} we replace one of the
logarithms by a derivative using
\begin{align}
  \label{repl2}
  \ln {k^2 \over k'^2} \, \left( \frac{k^2}{k'^2} \right)^{i \, \nu} =
  - i \frac{\pd}{\pd \nu} \left( \frac{k^2}{k'^2} \right)^{i \, \nu}
\end{align}
and integrate by parts. We arrive at
\begin{align}
  \label{NLOsol3}
  G(k,k', Y) = \int\limits_{-\infty}^{\infty} {d\nu \over 2\pi^2 \, k
    \, k'} \, e^{\left[\bar{\alpha}_\mu \, \chi_0(\nu) +
      \bar{\alpha}_\mu^2 \, \chi_1 (\nu)\right] \, Y} \, \left(
    \frac{k^2}{k'^2} \right)^{i \, \nu} \left( 1- \bam^2 \, \beta_2 \,
    \chi_0(\nu) \, Y \, \ln{k \, k'\over \mu^2}\right)
\end{align}
which, as can be easily verified, is $\mu$-independent up to order
${\cal O}(\bam^3)$
(cf. \cite{Ivanov:2005gn,Ivanov:2006gt,Caporale:2008is,Enberg:2005eq}). In
deriving \eq{NLOsol3}, and in the calculations to follow, we employ
the LLA power-counting: we assume that rapidity $Y$ is sufficiently
large, such that
\begin{align}
  \label{counting}
  \bam \, Y \sim 1,
\end{align}
and, with this assumption, only keep order-$\bam$ terms in the
prefactor. For instance, integration by parts performed in arriving at
\eqref{NLOsol3} also generates a term proportional to $\bam^3 \,
\chi'_1 (\nu) \, Y$, which is order-$\bam^2$ in our power-counting
and, being outside the precision of our NLO approximation, is
neglected. We only keep terms up to the order-$\bam$ both in the
exponent and in the prefactor in the NLO approximation.


\subsection{Searching for the NNLO BFKL Solution Ansatz}

Within the precision of our NLO power-counting (using \eq{counting} and
keeping only order-$\bam$ terms in the exponent and the prefactor) one
could rewrite \eq{NLOsol3} as
\begin{align}
  \label{NLOsol4}
  G(k,k', Y) = \int\limits_{-\infty}^{\infty} {d\nu \over 2\pi^2 \, k
    \, k'} \, e^{\left[\bas (k \, k') \, \chi_0(\nu) + \bas^2 (k \,
      k') \, \chi_1 (\nu)\right] \, Y} \, \left( \frac{k^2}{k'^2}
  \right)^{i \, \nu},
\end{align}
with the one-loop running coupling given by the expansion of
\eq{eq:beta2}.  Note that we can not uniquely fix the scale of the QCD
running coupling constant at this order: in fact, we can replace the
couplings in \eq{NLOsol4} by
\begin{align}
  \label{coup_repl}
  \bas (k \, k') \ \to \ \bas^\lambda (k^2) \ \bas^\lambda (k'^2) \
  \bas^{1 - 2 \, \lambda} (k \, k')
\end{align}
with any real number $\lambda$.\footnote{For instance, $\lambda = 1/2$
  leads to the square-root ansatz, $\sqrt{\bas (k^2) \ \bas (k'^2)}$,
  while $\lambda=1$ gives a "triumvirate" structure
  \cite{Braun:1994mw,Levin:1994di,Balitsky:2006wa,Kovchegov:2006vj}. The
  running of the coupling in \eq{NLOsol4} is recovered by putting
  $\lambda =0$.}  Perturbative expansions to order-$\bam^2$ are
equivalent on the left and right sides of \eqref{coup_repl}. To fix
the scale(s) of the couplings in \eq{NLOsol4} one has to
systematically extend the NLO BFKL solution presented above to NNLO.

The form of the solution of the NLO BFKL equation in \eq{NLOsol4} is
very appealing, since it suggests an elegant way to generalize the
solution of the all-order BFKL equation in ${\cal N} =4$ SYM theory to
QCD by simply replacing the fixed couplings of ${\cal N} =4$ SYM
theory by running QCD couplings. (Indeed the functions $\chi_i (\nu)$
are, in general, different in the two theories for $i \ge 1$
\cite{Kotikov:2000pm,Balitsky:2009xg,Balitsky:2009yp}.)

To that end, it appears natural to suggest and test the following
ansatz for the solution of the NNLO BFKL equation, which appears as a
straightforward extension of \eq{NLOsol4}:
\begin{align}
  \label{NNLOans1}
  G(k,k', Y) = \int\limits_{-\infty}^{\infty} {d\nu \over 2\pi^2 \, k
    \, k'} \, e^{\left[\bas (k \, k') \, \chi_0(\nu) + \bas^2 (k \,
      k') \, \chi_1 (\nu) + \bas^3 (k \,
      k') \, \chi_2 (\nu) \right] \, Y} \, \left( \frac{k^2}{k'^2}
  \right)^{i \, \nu}.
\end{align}
As mentioned above, $\chi_2 (\nu)$ is the $(\nu \leftrightarrow -
\nu)$-even (real) scale-invariant part of the prefactor generated by
the action of the NNLO BFKL kernel on the LO eigenvalue. The exact
form of $\chi_2 (\nu)$, while needed in \eq{NNLOans1}, is not required
to test whether the ansatz \eqref{NNLOans1} in fact does solve the
NNLO BFKL equation, just like we never used the exact form of $\chi_1
(\nu)$ in constructing the NLO BFKL solution \eqref{NLOsol}.

In constructing the NLO BFKL solution our knowledge of NLO kernel was
defined by \eq{conf-proj2}. The important to us NLO terms on its
right-hand-side were the term proportional to $\ln (k^2/\mu^2)$ and
the term containing $\chi'_0 (\nu)$. The latter term became complex
for $\gamma = \frac{1}{2} + i \, \nu$ and was canceled to give a real
eigenvalue \eqref{eigenv3}. The term proportional to $\ln (k^2/\mu^2)$
resulted from the running of the coupling in the LO BFKL eigenvalue:
it is clear that with $k$ being the only momentum scale left on the
right of \eq{conf-proj2} the coupling there has to run with the scale
$k^2$. One may conclude that the $k$-dependent terms appearing after
the action of the NNLO kernel on the LO eigenfunction are entirely due
to the running of the couplings in the LO and NLO terms with the scale
$k^2$. We can, therefore, write for the action of the NNLO kernel on
the LO eigenvalues as
\begin{align}
  \label{conf-proj3}
  \int d^2 q \, K^{{\rm LO}+ {\rm NLO}+ {\rm NNLO}}(k,q) \ q^{-1+2 i
    \nu} = \left\{ \bar{\alpha}_\mu \, \chi_0 (\nu) \left[ 1-
      \bar{\alpha}_\mu \, \beta_2 \, \ln{k^2\over \mu^2} + \bam^2
      \beta_2^2 \, \ln^2 {k^2\over \mu^2} + \bam^2 \, \beta_3 \,
      \ln{k^2\over \mu^2} \right] \right. \notag \\ \left. + \bam^2 \,
    \left[ \frac{i}{2} \, \beta_2 \, \chi'_0 (\nu) + \chi_1 (\nu)
    \right] \left[ 1- 2\, \bar{\alpha}_\mu \, \beta_2 \, \ln{k^2\over
        \mu^2} \right] + \bam^3 \, \left[ \chi_2 (\nu) + i \, \delta_2
      (\nu) \right] \right\} \, k^{-1 + 2 i \nu},
\end{align}
where 
\begin{align}
  \label{eq:LONLONNLO}
  K^{{\rm LO}+ {\rm NLO} + {\rm NNLO}}(k,q) \equiv \bar{\alpha}_\mu \,
  K^{\rm LO} (k,q) + \bar{\alpha}_\mu^2 \, K^{\rm NLO} (k,q) +
  \bar{\alpha}_\mu^3 \, K^{\rm NNLO} (k,q),
\end{align}
$\beta_3$ is the two-loop coefficient in the QCD beta-function
\cite{Caswell:1974gg,Jones:1974mm,Egorian:1978zx} defined by
\begin{align}
  \label{beta3}
  \mu^2 \frac{d \bam}{d \mu^2} = - \beta_2 \, \bam^2 + \beta_3 \,
  \bam^3,
\end{align}
and $i \, \delta_2 (\nu)$ denotes the $(\nu \leftrightarrow -
\nu)$-odd imaginary $k$-independent NNLO term on the right of
\eqref{conf-proj3}. 

While $\chi_2 (\nu)$ and $\delta_2 (\nu)$ are presently unknown, one
may still check whether the ansatz \eqref{NNLOans1} solves
\begin{align}
  \label{NNLO_BFKL}
  \partial_Y G (k, k', Y) = \int d^2 q \, K^{{\rm LO}+ {\rm NLO} +
    {\rm NNLO}} (k, q) \, G (q, k', Y)
\end{align}
up to order-$\bam^2$ in the power-counting of \eqref{counting}. In
Appendix B we show that this, in fact, is not the case, and the ansatz
\eqref{NNLOans1} does not solve the NNLO BFKL
equation.\footnote{Appendix B also shows explicitly that \eq{NLOsol4}
  does, in fact, solve the NLO BFKL equation.} The failure of the
ansatz \eqref{NNLOans1} appears to be unrelated to the choice of the
running of the coupling. Instead we propose a modified ansatz
\eqref{NNLOans2} for the solution of the NNLO BFKL equation, which
works only if $\delta_2 (\nu)$ is given by \eq{delta2}. Note that to
verify the ansatz \eqref{delta2}, \eqref{NNLOans2} one has to
construct NNLO eigenfunctions perturbatively, using the method
presented above for the NLO calculation \cite{GiovanniYuri2}.  (While
approximations for the NNLO BFKL eigenvalues exist in the literature
\cite{Ball:2005mj,Marzani:2007gk}, our solution procedure is not
designed to construct the conformal part $\chi_2 (\nu)$ of the NNLO
eigenvalue: hence construction of NNLO eigenfunctions using our method
would not allow one to derive the NNLO eigenvalues explicitly (except
for their imaginary parts) to be compared with
\cite{Ball:2005mj,Marzani:2007gk} or other works on the subject.)


\section{DGLAP Anomalous Dimension}
\label{sec:anomdim}

In Deep Inelastic Scattering (DIS) one is interested in the
Bjorken-$x$ dependence of the structure functions. It is different
from the energy-dependence in, say, the up-down symmetric
$\gamma^*\gamma^*$ scattering.  Indeed, in the case of the up-down
symmetric kernel like in the $\gamma^*\gamma^*$ scattering process,
the rapidity variable $Y$ that enters the BFKL equation
(\ref{eq:BFKL}) is $Y^{\rm sym}=\ln {s\over kk'}$ used through this
paper so far. In DIS we instead have $Y^{\rm DIS}=\ln {s\over k^2}
\approx \ln \frac{1}{x}$ where $s$ is the center-of-mass energy
squared of the virtual photon--hadron system, $k$ is the momentum of
the reggeized gluon in the BFKL ladder that is attached to the quark
anti-quark pair produced by the virtual photon. Assuming that $k
\approx Q$ with $Q$ the photon virtuality, we can identify $Y^{\rm
  DIS}$ with $\ln \frac{1}{x}$, where $x$ is the Bjorken scaling
variable \cite{Bjorken:1968dy}.

The modification needed to account for this difference in energy
dependence is irrelevant at LO but it becomes important at NLO
accuracy. As explained in \cite{Fadin:1998py}, in DIS case one has to
modify the evolution kernel by adding an extra term to it that
consequently introduces a new additive correction to the
``eigenvalues'' which is proportional to $\bam^2 \,
\chi_0(\gamma)\chi'_0(\gamma)$.

We will now show that our solution of the BFKL equation in the
symmetric case given by (\ref{NLOsol}) leads to the same term in the
eigenvalue for DIS: the term is generated by the change in the definition
of rapidity, without having to solve the eigenvalue problem again. In
this way one recovers exactly the term proportional to $\bam^2 \,
\chi_0(\gamma)\chi'_0(\gamma)$ that was found in \cite{Fadin:1998py}.

Writing $Y^{\rm sym} = Y^{\rm DIS} + \ln {k\over k'}$ in
(\ref{NLOsol}) yields
\begin{align}
  G \Bigg( k, k', & \, Y^{\rm sym} = Y^{\rm DIS} + \ln {k\over k'}
  \Bigg) \equiv {\tilde G} (k, k', Y^{\rm DIS}) \notag \\ &=
  \int\limits_{-\infty}^{\infty} {d\nu \over 2\pi^2} \,
  e^{\left[\bar{\alpha}_\mu \, \chi_0(\nu) + \bar{\alpha}_\mu^2 \,
      \chi_1 (\nu)\right] \, (Y^{\rm DIS}+ \ln {k\over k'}) } \,
  H_{\frac{1}{2}+i \, \nu} (k) \, \left[ H_{\frac{1}{2}+i \, \nu} (k')
  \right]^*
  \nonumber\\
  &\simeq \int\limits_{-\infty}^{\infty} {d\nu \over 2\pi^2} \,
  e^{\left[\bar{\alpha}_\mu \, \chi_0(\nu) + \bar{\alpha}_\mu^2 \,
      \chi_1 (\nu)\right] \, Y^{\rm DIS} } \, H_{\frac{1}{2}+i \, \nu}
  (k) \, \left[ H_{\frac{1}{2}+i \, \nu} (k') \right]^* \left(1+ \bam
    \, \chi_0(\nu) \, \ln {k\over k'}\right), \label{Gtilde}
\end{align}
where we have now put the rapidity-independent term into the
prefactor.  The NLO BFKL evolution equation in $Y^{\rm DIS}$ for the
Green function $\tilde G$ has the kernel \cite{Fadin:1998py}
\begin{align}\label{KDIS}
  K^{\rm LO+NLO} (k,q) - \int d^2 q' \, K^{\rm LO} (k, q') \, \ln
  \frac{q'}{k} \, K^{\rm LO} (q', q).
\end{align}
In general we need to find the eigenfunctions and eigenvalues of this
new kernel. However, it appears easier to simply try to cast
\eq{Gtilde} in the form where all the $k$ and $k'$-dependence is
contained in the product of a new eigenfunction and its complex
conjugate (with momentum $k'$). Since the new $k, k'$-dependent factor
in the parenthesis of the last line of \eq{Gtilde} is real and can not
be absorbed into the eigenfunctions, it has to be reabsorbed back into
the exponent. (Alternatively we can note that $H_{\frac{1}{2}+i \,
  \nu} (k)$ are, in fact, eigenfunctions of the kernel \eqref{KDIS}.)
To this end, with the NLO precision, we re-write $\ln {k\over k'}$ as
a $\nu$-derivative acting on $\left({k^2\over k'^2}\right)^{i\,\nu}$,
perform a partial integration and include the term proportional to
$Y^{\rm DIS}$ back into the exponential, obtaining
\begin{align}
 \label{NLO-DIS-sol}
 {\tilde G} (k, k', Y^{\rm DIS}) = \int\limits_{-\infty}^{\infty}
 {d\nu \over 2\pi^2} \, e^{\left[\bar{\alpha}_\mu \, \chi_0(\nu) +
     \bar{\alpha}_\mu^2 \, \big(\chi_1 (\nu) + \frac{i}{2}
     \,\chi_0(\nu) \, \chi'_0(\nu) \big)\right] \, Y^{\rm DIS} } \,
 H_{\frac{1}{2}+i \, \nu} (k) \, \left[ H_{\frac{1}{2}+i \, \nu} (k')
 \right]^* \left(1+ {i\over 2} \,\bam \,\chi'_0(\nu)\right).
\end{align}
It is clear from (\ref{NLO-DIS-sol}) that when this Green function
$\tilde G$ is either acted upon by the kernel \eqref{KDIS}, or simply
differentiated with respect to $Y^{\rm DIS}$, the new eigenvalues
\begin{align}
  \label{eigenv_DIS}
  \Delta^{\rm DIS} (\nu) = \bar{\alpha}_\mu \, \chi_0(\nu) +
  \bar{\alpha}_\mu^2 \, \left(\chi_1 (\nu) + \frac{i}{2} \,\chi_0(\nu)
    \, \chi'_0(\nu) \right)
\end{align}
obtained this way turn out to be shifted compared to \eqref{eigenv3}
by $\bam^2 \, \frac{i}{2} \,\chi_0(\nu) \, \chi'_0(\nu) = - \bam^2 \,
\frac{1}{2} \,\chi_0(\gamma) \, \chi'_0(\gamma)$, that is, by exactly
the same term that was obtained in \cite{Fadin:1998py}. Note that we
obtained this result without solving a new eigenvalue problem. In
other words, the $H_\gamma(k)$ functions are also eigenfunctions for
the non-symmetric NLO BFKL kernel of the DIS case. Notice also that
the eigenvalues are not anymore symmetric under $\nu\to-\nu$ because
now (\ref{NLO-DIS-sol}) is a solution of an evolution equation with a
non-symmetric kernel.

Expanding $\Delta^{\rm DIS} (\gamma)$ in the powers of $\gamma$ around
$\gamma = 0$ and solving $\omega = \Delta^{\rm DIS} (\gamma)$ equation
for $\gamma (\omega)$ one recovers the NNLO DGLAP anomalous dimension
in the small-$x_B$ limit, as was demonstrated in \cite{Fadin:1998py}.


\section{Conclusions and Outlook}
\label{sec:outlook}

In this paper we have derived the solution for the NLO BFKL equation
\eqref{NLOsol} by constructing its eigenfunctions using a perturbative
expansion around the LO eigenfunctions. This expansion procedure can
be used to construct solutions of the higher-order BFKL equation, as
suggested in \eq{all_sol} above, providing a way of organizing the QCD
perturbation series in the high energy scattering processes. It can
already be applied to study the NNLO BFKL equation
\cite{GiovanniYuri2}. Our solution method can also be applied to the
case of BFKL evolution with non-trivial azimuthal angle dependence,
and to the non-forward BFKL equation as well.

Other immediate applications of our result include a construction of
the NLO high-energy forward scattering amplitude for the $\gamma^*
\gamma^*$ scattering process.  The high-energy (Regge limit)
scattering amplitude of $\gamma^* \gamma^*$ process, which encodes the
hadronic contribution to, for example, the high-energy $e^+ e^-$
scattering, can be factorized into a convolution of the virtual
photons impact factors and the BFKL ladder (Green function)
exchange. Since the photon impact factors are known at NLO
\cite{Balitsky:2012bs,Balitsky:2010ze}, one can now use our NLO BFKL
Green function \eqref{NLOsol} to fully construct the $\gamma^*
\gamma^*$ forward scattering amplitude at NLO. This is also left for
future work \cite{GiovanniYuri2}.


\section*{Acknowledgments}

The authors are grateful to Ian Balitsky, Dima Kharzeev, and Al
Mueller for helpful discussions. This research is sponsored in part by
the U.S. Department of Energy under Grant No. DE-SC0004286. \\


\section*{Appendix A: Orthogonality}
\renewcommand{\theequation}{A\arabic{equation}}
  \setcounter{equation}{0}
\label{A}

We want to show that the functions $H_{\frac{1}{2} + i \, \nu} (k)$
from \eq{eigenf_nu2} satisfy the orthogonality relation in
\eq{NLO-ortho1}. The relation is clearly satisfied by the LO BFKL
kernel eigenfunctions. Substituting the eigenfunctions from
\eq{eigenf_nu2} into \eq{NLO-ortho1} we obtain at order-$\bam$
\begin{align}
  \label{ortho2}
  \int\limits_{-\infty}^\infty d\nu \, f(\nu) \, \int \frac{d^2
    k}{k^2} \, k^{2 \, i \, (\nu - \nu')} \, \Bigg\{ \frac{i \,
    \beta_2}{2} \, \left[ {\chi_0 (\nu) \over \chi'_0 (\nu)} - {\chi_0
      (\nu') \over \chi'_0 (\nu')} \right] \, \ln^2{k^2\over \mu^2} +
  \frac{\beta_2}{2} \left[ {\partial\over\partial\nu} \left( {\chi_0
        (\nu) \over \chi_0'(\nu)} \right) +
    {\partial\over\partial\nu'} \left( {\chi_0 (\nu') \over
        \chi_0'(\nu')} \right) \right] \ln{k^2\over \mu^2} \notag \\ +
  i \, \left( {\rm Im}[c_1(\nu)] - {\rm Im}[c_1(\nu')] \right)
  \ln{k^2\over \mu^2} + c_0(\nu) + c_0^* (\nu') \Bigg\} =0
\end{align}
where $f(\nu)$ is an arbitrary test function. In the first term in
\eqref{ortho2} we rewrite one power of the logarithm using
\begin{align}
  \label{ortho3}
  \left( \ln{k^2\over \mu^2} \right) \, k^{2 \, i \, (\nu - \nu')} =
  \left( - i \frac{\pd}{\pd \nu} - \ln \mu^2 \right) \, k^{2 \, i \,
    (\nu - \nu')}
\end{align}
and integrate by parts over $\nu$ to get
\begin{align}
  \label{ortho4}
  \int\limits_{-\infty}^\infty d\nu \, \int \frac{d^2 k}{k^2} \, k^{2
    \, i \, (\nu - \nu')} \, \Bigg\{ - \frac{\beta_2}{2} \,
  \frac{\pd}{\pd \nu} \left( {\chi_0 (\nu) \over \chi'_0 (\nu)}
  \right) \, f(\nu) \, \ln{k^2\over \mu^2} - \frac{\beta_2}{2} \,
  \left[ {\chi_0 (\nu) \over \chi'_0 (\nu)} - {\chi_0 (\nu') \over
      \chi'_0 (\nu')} \right] \, f'(\nu) \, \ln{k^2\over \mu^2} \notag
  \\ - \frac{i \, \beta_2}{2} \, \left[ {\chi_0 (\nu) \over \chi'_0
      (\nu)} - {\chi_0 (\nu') \over \chi'_0 (\nu')} \right] \, f(\nu)
  \, \ln{k^2\over \mu^2} \, \ln{\mu^2} + \frac{\beta_2}{2} \left[
    {\partial\over\partial\nu} \left( {\chi_0 (\nu) \over
        \chi_0'(\nu)} \right) + {\partial\over\partial\nu'} \left(
      {\chi_0 (\nu') \over \chi_0'(\nu')} \right) \right] \, f(\nu) \,
  \ln{k^2\over \mu^2} \notag \\ + i \, \left( {\rm Im}[c_1(\nu)] -
    {\rm Im}[c_1(\nu')] \right) \, f(\nu) \, \ln\frac{k^2}{\mu^2} + 2
  \, {\rm Re} [c_0(\nu)] \, f(\nu) \Bigg\} =0.
\end{align}
Canceling the first term with the first part of the fourth term, and
applying the substitution \eqref{ortho3} to all remaining logarithms
of $k^2$ yields after a few more cancellations
\begin{align}
  \label{ortho5}
  \int\limits_{-\infty}^\infty d\nu \, \int \frac{d^2 k}{k^2} \, k^{2
    \, i \, (\nu - \nu')} \, \Bigg\{ - \frac{i \, \beta_2}{2} \,
  \left[ {\chi_0 (\nu) \over \chi'_0 (\nu)} - {\chi_0 (\nu') \over
      \chi'_0 (\nu')} \right] \, f''(\nu) + \frac{\beta_2}{2} \,
  \left[ {\chi_0 (\nu) \over \chi'_0 (\nu)} - {\chi_0 (\nu') \over
      \chi'_0 (\nu')} \right] \, f' (\nu) \, \ln \mu^2 \notag \\ -
  \left( {\rm Im}[c_1(\nu)] - {\rm Im}[c_1(\nu')] \right) \, f'(\nu)
  \Bigg\} =0,
\end{align}
where we have also employed \eq{cond2}. Integrating over $k$ gives a
delta-function $\delta (\nu - \nu')$, which makes the expression in
the curly brackets vanish, thus showing that \eq{ortho5} is an
identity. 

We have, therefore, shown that the functions $H_{\frac{1}{2} + i \,
  \nu} (k)$ from \eq{eigenf_nu2} satisfy the orthogonality relation in
\eq{NLO-ortho1} up to order-$\bam$, which is the limit of precision of
the NLO approximation.


\section*{Appendix B: Verifying the NNLO BFKL Solution Ansatz}
\renewcommand{\theequation}{B\arabic{equation}}
  \setcounter{equation}{0}
\label{B}

The goal of this Appendix is to verify whether the ansatz
\eqref{NNLOans1} solves \eq{NNLO_BFKL}. (The initial condition
\eqref{eq:init} is trivially satisfied by the ansatz.) Start with the
right-hand-side of the equation \eqref{NNLO_BFKL}: substituting the
ansatz \eqref{NNLOans1}, replacing
\begin{align}
  \label{repl3}
  \ln \frac{k \, k'}{\mu^2} \ \to \ \ln \frac{k'^2}{\mu^2} -
  \frac{i}{2} \, \frac{\pd}{\pd \nu}
\end{align}
in the arguments of the running couplings in the exponent with the
derivative acing only on $(k^2/k'^2)^{i \, \nu}$, and employing
\eq{conf-proj3} we get
\begin{align}
  \label{rhs1}
  & \int d^2 q \ K^{{\rm LO}+ {\rm NLO} + {\rm NNLO}} (k, q) \, G (q,
  k', Y) = \int\limits_{-\infty}^{\infty} {d\nu \over 2\pi^2 \, k \,
    k'} \notag \\ & \times \, \exp \left\{ \bam \, \chi_0(\nu) \, Y \,
    \left[ 1- \bar{\alpha}_\mu \, \beta_2 \, \left( \ln
        \frac{k'^2}{\mu^2} - \frac{i}{2} \, \frac{\pd}{\pd \nu}
      \right) + \bam^2 \beta_2^2 \, \left( \ln \frac{k'^2}{\mu^2} -
        \frac{i}{2} \, \frac{\pd}{\pd \nu} \right)^2 + \bam^2 \,
      \beta_3 \, \left( \ln \frac{k'^2}{\mu^2} - \frac{i}{2} \,
        \frac{\pd}{\pd \nu} \right) \right] \right. \notag \\ &
  \left. + \bam^2 \, \chi_1 (\nu) \, Y \, \left[ 1- 2 \,
      \bar{\alpha}_\mu \, \beta_2 \, \left( \ln \frac{k'^2}{\mu^2} -
        \frac{i}{2} \, \frac{\pd}{\pd \nu} \right)\right] + \bam^3 \,
    \chi_2 (\nu) \, Y \right\} \notag \\ & \times \, \Bigg\{
  \bar{\alpha}_\mu \, \chi_0 (\nu) \left[ 1- \bar{\alpha}_\mu \,
    \beta_2 \, \ln{k^2\over \mu^2} + \bam^2 \beta_2^2 \, \ln^2
    {k^2\over \mu^2} + \bam^2 \, \beta_3 \, \ln{k^2\over \mu^2}
  \right] + \bam^2 \, \left[ \frac{i}{2} \, \beta_2 \, \chi'_0 (\nu) +
    \chi_1 (\nu) \right] \left[ 1- 2\, \bar{\alpha}_\mu \, \beta_2 \,
    \ln{k^2\over \mu^2} \right] \notag \\ & + \bam^3 \, \left[ \chi_2
    (\nu) + i \, \delta_2 (\nu) \right] \Bigg\} \left(
    \frac{k^2}{k'^2} \right)^{i \, \nu},
\end{align}
where now the $\nu$-derivatives act on everything to the right of the
exponential. After a considerable algebra, involving numerous
differentiations and integrations by parts, and expanding the
exponential in the powers of running-coupling terms (that is, the
terms containing $\beta_2$ and $\beta_3$) while keeping terms up to
order-$\bam^3$ (in the power-counting of \eqref{counting}) in the
expression, we obtain for the right-hand-side
\begin{align}
  \label{rhs2}
  & \int d^2 q \ K^{{\rm LO}+ {\rm NLO} + {\rm NNLO}} (k, q) \, G (q,
  k', Y) = \notag \\ & \int\limits_{-\infty}^{\infty} {d\nu \over
    2\pi^2 \, k \, k'} \, \left( \frac{k^2}{k'^2} \right)^{i \, \nu}
  \, e^{\left[ \bam \, \chi_0 (\nu) + \bam^2 \, \chi_1 (\nu)
      +\bam^3\,\chi_2(\nu)\right] \, Y} \, \Bigg\{ \bam \, \chi_0
  (\nu) \, \left[ 1 - \bam \, \beta_2 \, \ln \frac{k \, k'}{\mu^2} +
    (\bam \, \beta_2)^2 \, \left( \ln^2 \frac{k \, k'}{\mu^2} +
      \frac{\chi''_0 (\nu)}{4 \, \chi_0 (\nu)} \right) \right] \notag
  \\ & - [\bam \, \chi_0 (\nu)]^2 \, Y \, \left[ \bam \, \beta_2 \,
    \ln \frac{k \, k'}{\mu^2} - 2 \, (\bam \, \beta_2)^2 \, \left(
      \ln^2 \frac{k \, k'}{\mu^2} - \frac{\chi''_0 (\nu)}{4 \, \chi_0
        (\nu)} + \frac{\chi'_0 (\nu)^2}{8 \, \chi_0 (\nu)^2} \right)
  \right] + \bam^2 \, \chi_1 (\nu) \, \left[ 1 - 2\, \bam \, \beta_2
    \, \ln \frac{k \, k'}{\mu^2} \right] \notag \\ & + \frac{1}{2} \,
  [\bam \, \chi_0 (\nu)]^3 \, Y^2 \, (\bam \, \beta_2)^2 \, \left(
    \ln^2 \frac{k \, k'}{\mu^2} - \frac{\chi''_0 (\nu)}{4 \, \chi_0
      (\nu)} \right) + \bam^3 \, \chi_0 (\nu) \, \beta_3 \, \left( 1 +
    \bam \, \chi_0 (\nu) \, Y \right) \, \ln \frac{k \, k'}{\mu^2}
  \notag \\ & - 3 \, \bam^3 \, \chi_0 (\nu) \, \chi_1 (\nu) \, Y \,
  \bam \, \beta_2 \, \ln \frac{k \, k'}{\mu^2} + \frac{i}{2} \,
  \chi'_0 (\nu) \, \bam^3 \, \beta_3 - i \chi'_1 (\nu) \, \bam^3 \,
  \beta_2 + \bam^3 \, \left[ \chi_2 (\nu) + i \, \delta_2 (\nu)
  \right] \Bigg\}.
\end{align}

After a somewhat lesser amount of calculation involving a similar
expansion of the exponential in the running-coupling terms we obtain
for the left-hand-side of \eq{NNLO_BFKL}
\begin{align}
  \label{lhs1}
  \pd_Y \, G (k, k', Y) & = \int\limits_{-\infty}^{\infty} {d\nu \over
    2\pi^2 \, k \, k'} \, \left( \frac{k^2}{k'^2} \right)^{i \, \nu}
  \, e^{\left[ \bam \, \chi_0 (\nu) + \bam^2 \, \chi_1 (\nu)
      +\bam^3\,\chi_2(\nu)\right] \, Y} \, \Bigg\{ \bam \, \chi_0
  (\nu) \, \left[ 1 - \bam \, \beta_2 \, \ln \frac{k \, k'}{\mu^2} +
    (\bam \, \beta_2)^2 \, \ln^2 \frac{k \, k'}{\mu^2} \right] \notag
  \\ & - [\bam \, \chi_0 (\nu)]^2 \, Y \, \left[ \bam \, \beta_2 \,
    \ln \frac{k \, k'}{\mu^2} - 2 \, (\bam \, \beta_2)^2 \, \ln^2
    \frac{k \, k'}{\mu^2} \right] + \bam^2 \, \chi_1 (\nu) \, \left[ 1
    - 2\, \bam \, \beta_2 \, \ln \frac{k \, k'}{\mu^2} \right] \notag
  \\ & + \frac{1}{2} \, [\bam \, \chi_0 (\nu)]^3 \, Y^2 \, (\bam \,
  \beta_2)^2 \, \ln^2 \frac{k \, k'}{\mu^2} + \bam^3 \, \chi_0 (\nu)
  \, \beta_3 \, \left( 1 + \bam \, \chi_0 (\nu) \, Y \right) \, \ln
  \frac{k \, k'}{\mu^2} \notag \\ & - 3 \, \bam^3 \, \chi_0 (\nu) \,
  \chi_1 (\nu) \, Y \, \bam \, \beta_2 \, \ln \frac{k \, k'}{\mu^2} +
  \bam^3 \, \chi_2 (\nu) \Bigg\}.
\end{align}

Comparing the left-hand-side \eqref{lhs1} with the right-hand-side
\eqref{rhs2} we see a complete agreement to order-$\bam^2$ (in the
power-counting of \eqref{counting}). This provides an independent
cross-check that \eq{NLOsol4} is, in fact, a solution of the NLO BFKL
equation.

However, at order-$\bam^3$ the expressions in Eqs.~\eqref{lhs1} and
\eqref{rhs2} are not equal to each other. In fact, the difference is
non-zero and is equal to
\begin{align}
  \label{diff1}
  \pd_Y \, G (k, k', Y) - \int d^2 q \ & K^{{\rm LO}+ {\rm NLO} + {\rm
      NNLO}} (k, q) \, G (q, k', Y) = - \int\limits_{-\infty}^{\infty}
  {d\nu \over 2\pi^2 \, k \, k'} \, \left( \frac{k^2}{k'^2} \right)^{i
    \, \nu} \, e^{\left[ \bam \, \chi_0 (\nu) + \bam^2 \, \chi_1 (\nu)
      +\bam^3\,\chi_2(\nu)\right] \, Y} \notag \\ & \times \, \Bigg\{
  \bam^3 \, \beta_2^2 \, \Bigg[ \frac{\chi''_0 (\nu)}{4} + \bam \,
  \chi_0 (\nu) \, Y \, \left( - \frac{\chi''_0 (\nu)}{2} +
    \frac{\chi'_0 (\nu)^2}{4 \, \chi_0 (\nu)} \right) - \left[ \bam \,
    \chi_0 (\nu) \, Y \right]^2 \, \frac{\chi''_0 (\nu)}{8} \Bigg]
  \notag \\ & + \frac{i}{2} \, \chi'_0 (\nu) \, \bam^3 \, \beta_3 - i
  \chi'_1 (\nu) \, \bam^3 \, \beta_2 + \bam^3 \, i \, \delta_2 (\nu)
  \Bigg\}.
\end{align}
Hence the ansatz \eqref{NNLOans1} can not be a solution of the NNLO
BFKL equation, independent of what $\delta_2 (\nu)$ is equal to. 

Using \eq{diff1} we can correct the ansatz \eqref{NNLOans1}. First we
notice that requiring the imaginary terms on the right of \eq{diff1}
to separately vanish fixes $\delta_2 (\nu)$ to be given by
\begin{align}
  \label{delta2}
  \delta_2 (\nu) = - \frac{1}{2} \, \chi'_0 (\nu) \, \beta_3 + \chi'_1
  (\nu) \, \beta_2.
\end{align}
Note that a complete perturbative construction of NNLO
eigenfunctions, together with hermiticity of the NNLO BFKL kernel,
would uniquely fix $\delta_2 (\nu)$: without doing that we may only
conjecture that $\delta_2 (\nu)$ is given by the expression
\eqref{delta2}.

Finally, for $\delta_2 (\nu)$ from \eq{delta2}, we can correct the
ansatz \eqref{NNLOans1} by a multiplicative factor in the integrand,
obtained by adding to unity the integral over $Y$ of the real term in
the curly brackets on the right-hand-side of \eq{diff1}: 
\begin{align}
  \label{NNLOans2}
  & G(k,k', Y) = \int\limits_{-\infty}^{\infty} {d\nu \over 2\pi^2 \,
    k \, k'} \, e^{\left[\bas (k \, k') \, \chi_0(\nu) + \bas^2 (k \,
      k') \, \chi_1 (\nu) + \bas^3 (k \, k') \, \chi_2 (\nu) \right]
    \, Y} \, \left( \frac{k^2}{k'^2} \right)^{i \, \nu} \notag \\
  & \times \, \left\{ 1 + (\bam \, \beta_2)^2 \, \left[ - \frac{1}{24}
      \, (\bam \, Y)^3 \, \chi_0 (\nu)^2 \, \chi''_0 (\nu) +
      \frac{1}{4} \, (\bam \, Y)^2 \, \chi_0 (\nu) \, \left(
        \frac{\chi'_0 (\nu)^2}{2 \, \chi_0 (\nu)} - \chi''_0 (\nu)
      \right) + \bam \, Y \, \frac{\chi''_0 (\nu)}{4} \right]
  \right\}.
\end{align}

Let us stress once more that to verify the ansatz of
Eqs.~\eqref{NNLOans2} and \eqref{delta2}, and find the true value of
$\delta_2 (\nu)$ one has to construct NNLO eigenfunctions
perturbatively with the help of \eq{conf-proj3}, similar to how NLO
eigenfunctions were found above. This is beyond the scope of the
current work and is left for the future \cite{GiovanniYuri2}.



\providecommand{\href}[2]{#2}\begingroup\raggedright\endgroup

\end{document}